\documentstyle[12pt]{article}
\begin{document}

\begin{titlepage}

{\hspace{9cm}{\vbox{\hbox{hep-th/9508073}
\hbox{SLAC-PUB-95-6986}\hbox{1995}}}}

\vspace{1cm}
{\LARGE
{\vbox{\centerline{Grand Unification with Three
 Generations}\smallskip
\centerline{in Free Fermionic String Models }}}
}
\vspace{1.5cm}{\Large
\centerline{D. Finnell$^1$}
\vspace{.5cm}\centerline{Stanford Linear Accelerator Center}
\centerline{Stanford University, Stanford, California 94309, USA}}
\bigskip\bigskip\bigskip
\centerline{\bf ABSTRACT}
\bigskip\bigskip

\noindent
We examine the problem of constructing three generation free
fermionic string
models with grand unified gauge groups.  We attempt the construction
of $G\times G$ models, where $G$ is a grand unified group
realized at level 1.  This structure allows those Higgs
representations to appear which are necessary to break the
symmetry down to the standard model gauge group. For $G=SO(10)$,
we find only models with an even number of generations.  However,
for $G=SU(5)$ we find a number of 3 generation models.
\vfill

{\small\noindent
$^1$Address after Sept. 1, 1995: Deptartment of Physics and Astronomy,
Rutgers University, Piscataway NJ 08855-0849.}

\end{titlepage}
\section{Introduction}

The apparent unification of the gauge couplings of the
MSSM \cite{couplings} is a
successful prediction of supersymmetric grand unified theories.
In string theory,
the unification of coupling constants is also predicted\cite{ginsparg},
although at a
scale about one order of
magnitude higher than the $3\cdot10^{16}\mbox{GeV}$ indicated by
experiment\cite{vadim}. This discrepancy may be the result of
threshold corrections or charged matter at an intermediate scale
 modifying the naive
prediction (for a recent review see \cite{dienes}), or a nonstandard
normalization of the $U(1)$ hypercharge \cite{hyper}.
It is also possible that the string theory leads to a grand unified
group which  is spontaneously broken below the string scale.
This has motivated a search for realistic string models
featuring grand unified (GUT) groups.

It has been known for some time that the incorporation of grand
unification within string theory poses challenges.  The most basic
representations of gauge symmetries on the string world sheet, the
$k=1$ representations, do not allow for scalars in the adjoint
representation, necessary for the breaking of GUT symmetry, at the same
time as chiral fermions\cite{dkv,dlnr,lewellen}.
This problem can be overcome by
the construction of string models with gauge groups realized at
$k\ge2$.   Although such constructions are technically complicated,
models have been constructed both in the orbifold and free-fermion
formulations which feature the basic ingredients of a realistic GUT
group ($SO(10)$ or $SU(5)$) adjoint Higgs scalars and chiral
fermions\cite{lewellen,chung,newfont}.

One feature which has proven difficult to obtain in these models is the
correct three-fold replication of chiral fermion families. To date, a
variety of non-unified or semi-unified level 1
three generation string models have been constructed
in the literature \cite{yau,flipped,alr,froggie,cls,shamit}.
Methods for constructing
such models in the free-fermion or orbifold frameworks are well
established.  However, despite effort by a number of
groups \cite{chung,newfont,cleaver,volkov}, there exist no examples in the
literature of consistent string GUT models with three generations.

In this paper, we will focus on the specific problem of constructing a
string model which combines grand unification with three generations of
chiral fermions. We will approach the problem from a practical point of
view, attempting explicit constructions using the free-fermionic
formulation for string compactifications. Rather than constructing
$k=2$ models, we will follow the suggestion of
\cite{font,barbieri,volkov} and construct models
which contain a direct product of gauge factors $G_i\ge SU(5)$, each
realized at level 1, with massless scalars transforming as vectors
simultaneously under two groups. Vacuum expectation values (VEV's) for
these fields can then break the gauge group down to the standard model
embedded in a diagonal subgroup. For example, we will attempt to build
$SO(10)\times SO(10)$ models with
a $(\mbox{${\bf 10}$},\mbox{${\bf 10}$})$ representation,
or $SU(5)\times SU(5)$ with
$(\mbox{${\bf 5}$},\mbox{${\bf\bar5}$})+(\mbox{${\bf\bar5}$},
\mbox{${\bf 5}$})$. A VEV along the D-flat direction
\begin{equation}
\langle H_{i\bar{j}}\rangle=\langle \bar{H}_{\bar{i}j}\rangle=
{\rm diag}(X,X,X,Y,Y).\end{equation}
with $X\simeq Y\simeq M_{{\rm GUT}}$, $Y\ne X$, would break $SU(5)\times
SU(5)$ down to the diagonal subgroup $(SU(3)\times SU(2) \times
U(1))_D$, and the three standard model gauge couplings would be unified
at $X$, even if the couplings of the two $SU(5)$ factors were not
equal.  This mimics a conventional GUT group, with the mixed
representations $H,\bar{H}$ playing the role of the adjoint Higgs.
For these reasons we will refer to these models as ``unified''
models even though the GUT group is not simple.  The
fact that such models can be built with level 1 representations allows
us to avoid much of the technical baggage associated with higher level
models and concentrate on obtaining three generations.  Sting models
featuring this construction have been investigated extensively by
Maslikov et. al. \cite{volkov}.

We will organize our discussion as follows.  In Sections 2 and 3, we
will review the technique employed  in constructing level 1
three generation free fermion models, constructing
$SO(10)$ and $SU(5)$ models without Higgs scalars as a demonstration.
In Section 4, we will apply this technique to
models with the enlarged gauge sector
$SO(10)\times SO(10)$ with $(\mbox{${\bf 10}$},\mbox{${\bf 10}$})$ Higgs
representations, and study the complications
which arise.  Finally, in Section 5, we will identify
a construction which allows us to obtain a three generation
spectrum in $SU(5)\times SU(5)$ models, and we will present
a number of explicit models.

\section{Free Fermion Conventions}

In free fermion models, the string degrees of freedom needed to
cancel the conformal anomaly consist of 20 real right moving
(10 complex) and 44 real (22 complex) left moving free
fermions.
The rules for the construction of consistent free fermion
string  models were worked out in \cite{klt,abk}.
We will follow the conventions of \cite{klt},
where the boundary conditions imposed upon free fermions under
parallel transport about noncontractable loops are expressed
in terms of vectors $V$ such that
$\Psi^l\rightarrow -e^{-2\pi i V^l}
\Psi^l$,
where $-{1\over 2}\le  V^l_i   < {1\over 2}$.
The boundary conditions are generated by a set of
basis vectors $V_i$.  Let $m_i$ be the smallest integer for which
$m_iV_i=0\ (\mbox{mod\ 1})$.  Then the basis vectors must satisfy
\begin{eqnarray}
m_i V_i\cdot V_j = m_j V_i\cdot V_j&=&0 \quad (\mbox{mod\ 1})\\
2 m_i V_i\cdot V_i&=&0\quad  (\mbox{mod\ 1}),
\label{wconstraint}\end{eqnarray}
where
dot products are defined with Lorentzian signature
\begin{equation}
V_i\cdot V_j =-\sum_{{\rm right}}\delta_l
V_i^l V_j^l + \sum_{{\rm left}}\delta_l V_i^l V_j^l
\label{dot}\end{equation}
($\delta_l=1/2$ for a real fermion, $1$ for a complex one).
Phases appearing in the partition function are
defined in terms of an auxiliary matrix
$k_{ij}$, satisfying
\begin{eqnarray}
k_{ij}+k_{ji}&=&V_i\cdot V_j \quad (\mbox{mod\ 1}) \\
m_j k_{ij}&=&0 \quad (\mbox{mod\ 1})\\
k_{ii}+k_{i0} + s_i -{1\over 2}V_i\cdot V_i&=&0 \quad (\mbox{mod\ 1})
\label{kconstraint}\end{eqnarray}
Note that $s_i$ represents the boundary condition for the spacetime
free fermion.
To specify a string model, one must specify both the set of
basis vectors and the $k$-matrix.  We will follow the
convention of stating the nonvanishing lower half-diagonal
$k$-matrix elements, allowing the full $k$-matrix to be reconstructed
using the above equations.

Every set of boundary conditions generated by linear combinations
of the basis vectors defines a different sector
of the Hilbert space of string states.
States to be identified with physical particles
must satisfy
$m^2=H^L=H^R$, where $H^L,H^R$ are the left and right
moving worldsheet Hamiltonians.
The vacuum energy of $H^L$ is
\begin{equation}
E_0^{L,V}=-{1\over 12}+\sum_{l,{\rm left}}
{1\over 2}((V^l)^2-{1\over12})
\delta_l.\label{lenergy}
\end{equation}
and a similar equation holds for $E_R$.
Physical states must also satisfy a generalized GSO projection
for each basis vector:
\begin{equation}
V_i\cdot N_{\alpha V}= \sum_j k_{ij}\alpha_j+s_i-V_i\cdot
\overline{\alpha V} \quad (\mbox{mod\ 1})
\label{projection}\end{equation}
where $N_{\alpha V}$ is the vector of fermion excitations in the
sector $V=\sum_i \alpha_i V_i$.

Following standard notation, taking a ``$1$'' to represent
$V=-1/2$, we take
\begin{displaymath}
V_0=(1^{20} | 1^{44})
\end{displaymath}
as our first basis  vector. We will choose as our second basis vector
the conventional choice for the gravitino sector
\begin{displaymath}
V_1=(1^2 (100)^6 | 0^{44}).
\end{displaymath}
We choose the standard form of the worldsheet supercurrent
\begin{equation}
T_F=\bar\psi^\mu\partial X_\mu + i\sum_I  \bar\chi_I \bar{y}_I
\bar\omega_I.\label{supercurrent}
\end{equation}
which must also be preserved by boundary conditions. We will
use only periodic and antiperiodic boundary conditions for
the supersymmetric side of the string, for which the
``triplet constraint''\cite{klt}
\begin{equation}
V_j(\bar\chi_i)+V_j(\bar{y}_i)+V_j(\bar\omega_i)=
s_j \quad (\mbox{mod\ 1}),
\label{triplet}\end{equation}
is sufficient to guarantee superconformal invariance.

\section{The standard mechanism for three generations}

The three generation free fermion models existing in the
literature are based on a set of basis vectors
known as
the ``NAHE'' set\cite{flipped,froggie}:
\begin{eqnarray}
V_2&=&(1_c (100)^2 (010)^2 (010)^2|
(00)^2(10)^2 (10)^2 1^{10} (100)_c 0^{16})\nonumber\\
V_3&=&(1_c (010)^2 (100)^2 (001)^2|(10)^2(00)^2(01)^2
1^{10} (010)_c 0^{16})\nonumber\\
V_4&=&(1_c (001)^2(001)^2(100)^2|(01)^2(01)^2(00)^2 1^{10}
(001)_c 0^{16})
\end{eqnarray}
where we have used the notation
\begin{displaymath}
V_i=(S_i (\bar\chi_1 \bar{y}_1 \bar\omega_1)\ldots
|(y_1\omega_1)\ldots
\Psi_a^{10} (\eta_1\eta_2\eta_3)_c
\Psi_b^{16}).
\end{displaymath}
The subscript ``$c$'' denotes boundary conditions for complex, as
opposed to real, fermions. The group of ten fermions with identical
boundary conditions, $\Psi^{10}_a$, leads to an $SO(10)$ gauge group,
while $\Psi^{16}_b$ generate an $SO(16)$ group (which may be enhanced
to $E_8$ by gauge bosons from twisted sectors).   Also note, at this
stage, each $\eta_i$ will pair up with some $y_i$'s and $\omega_i$'s to
form an $SO(6)^3$ horizontal gauge group. Thus the total gauge group is
$SO(10)\times E_8 \times SO(6)^3$. This set of basis vectors has
been used as the starting point for standard-like and ``flipped''
$SU(5)$ models; the origin of three generations in these models
is studied in \cite{nanof,orbifrog}.
It will be instructive to ignore for the moment the complications which
follow from the need for adjoint Higgs scalars and see what
is involved in building a three generation $SO(10)$ (or
$SU(5)$) model based on the above basis vectors.

The above three basis vectors represent the three distinct classes of
right-hand-side boundary conditions one can have, which correspond to
the three twists of the $Z_2\times Z_2$ orbifold \cite{orbifrog} $(+ -
-)$,$(- + -)$, and $(- - +)$ respectively, acting on three 2-tori. This
set of twists breaks the $N=4$ supersymmetry to $N=1$.
In the standard approach, these three
distinct twists are to be the origin of three generations, one
generation arising from each.
Each of the above sectors, $V_{2,3,4}$, has been
chosen to give massless chiral fermions in the ${\bf 16}$ representation
of $SO(10)$. Take sector $V_2$ for example.  The vacuum energy is
$(E_R,E_L)=(0,0)$, so unexcited states from this sector are massless.
The fermion zero modes of $\Psi_a^{10}$ realize
spinor representations of $SO(10)$ (${\bf 16}$'s
and ${\bf \overline{16}}$'s). However,
there are additional zero modes, the zero modes of $S_i$,
$\bar\chi_{1,2}$, $\bar{y}_{3-6}$, $y_{3-6}$, and $\eta_1$.
Prior to the GSO projections, the original Clifford vacuum has
dimension $2^{11}=2^6\cdot{\bf 16}+2^6\cdot{\bf \overline{16}}$
for each spacetime
chirality. We can group the anticommuting fermion zero modes into sets of
commuting operators, $\bar{\Gamma}^0=S^\dagger S$,
$\Gamma^{10}=i\prod_a \Psi^{10}_{a}$,
$i\bar{\chi}_1\bar{\chi}_2$, $i\bar{y}_Iy_I$, and
$\eta_1^{\dagger}\eta_1$, whose eigenvalues will label the
degenerate states. Modulo 1, the eigenvalues of these are related to
fermion excitation numbers appearing in Eq. (\ref{projection})
by \cite{klt}
 \begin{equation} {1\over 2}(1-\Gamma_i)=N_i\quad \mbox{mod\ 1}. \end{equation}
The eigenvalue of $\Gamma^{10}$ determines the $SO(10)$ chirality
(${\bf 16}$ vs. ${\bf \overline{16}}$), while that of
$\bar\Gamma^0$ determines the
spacetime chirality. The GSO projection due to $V_3$ (or
$V_4$) (Eq. (\ref{projection}))
correlates $SO(10)$ chirality with spacetime chirality for
massless states; only left-handed ${\bf 16}$'s and
right handed ${\bf \overline{16}}$'s survive from this sector.
The value of $i\bar\chi_1\bar\chi_2$ is also
tied to spacetime chirality. We are left with the five internal fermion
operators $i\bar{y}_iy_i$, $i=3\ldots 6$, and $\eta_1^{\dagger}
\eta_1$, satisfying the constraint $\eta^\dagger\eta\prod_i
(i\bar{y}_iy_i)=+1$. This leaves an overall degeneracy of
$2^{5-1}=16$ for every chiral ${\bf 16}$.  They can be shown to
transform as two copies of ${\bf 4}+{\bf\bar{4}}$
under the $SO(6)$ horizontal groups.

At this stage, the model has a total of 48 generations. To reduce this
number,  one must add additional basis vectors, such that only one
replication of ${\bf 16}$ survives GSO projections in each of the above
sectors. In choosing additional basis vectors, it is simplest to
preserve the pairing between left and right moving fermions,
$V(\bar{y}^I)=V(y^I)$, $V(\bar\omega^I) =V(\omega^I)$. We could
alternately choose
boundary conditions which broke this pairing, instead pairing on the
left and right sides separately giving models with complex
fermions\footnote{A third choice \cite{lewellen,chung}, taking
fermions with unpairably real boundary conditions, will not be used
here.}.
Because of the Lorentzian signature of the dot product, Eq.
\ref{dot}, the left-right paired basis vectors automatically satisfy
Eqs. \ref{wconstraint}.  We may choose, for example:
\begin{eqnarray*}
V_5&=&(0_c\ 011\ 000\ 000\ 000\ 000\ 011 |
11\ 00\ 00\ 00\ 00\ 11\ 0^{10} 0^3_c 0^{16}) \\
V_6&=& (0_c\ 000\ 011\ 011\ 000\ 000\ 000 |
00\ 11\ 11\ 00\ 00\ 00\
0^{10} 0^3_c 0^{16} ) \\
V_7&=&(0_c\ 000\ 000\ 000\ 011\ 011\ 000 |
00\ 00\ 00\ 11\ 11\ 00\
0^{10} 0^3_c 0^{16}).
\end{eqnarray*}
This set of twists in the internal fermions breaks the horizontal
symmetry $SO(6)^3\rightarrow U(1)^6$;
the six remaining $U(1)$ currents are
$\eta_i^{\dagger} \eta_i$, and three currents associated with the
internal fermions, $J_1=iy_4y_5, J_2=iy_1\omega_6,
J_3=i\omega_2\omega_3$. There is now only a
two-fold degeneracy, corresponding to the two different eigenvalues of
the current $iy_4y_5=\pm1$.  $V_3$ and $V_4$ are similar, each having a
${\bf 16}_\pm$ under their corresponding $U(1)$'s. This model now has 6
generations. Note that, with the addition of these basis vectors, new
sectors containing massless nonchiral
 ${\bf 16}+{\bf \overline{16}}$ pairs have
appeared, for example $V_2+V_7$, $V_3+V_5$ and $V_4+V_6$.

To make the final reduction to three generations, we can attempt
to add another left-right symmetric vector which breaks  down the
internal $U(1)$'s labeling the degenerate ${\bf 16}$'s.  Such a
vector is, for example:
\begin{displaymath}
V_8=(0_c\ 011\ 000\ 011\ 000\ 011\ 000 |
11\ 00\ 11\ 00\ 11\ 00\
0^{10} 0^3_c 0^{16}).
\end{displaymath}
This vector indeed leaves only one ${\bf 16}$ from each of the three
sectors $V_2,V_3,V_4$. However, we find the unpleasant feature that new,
unwanted chiral ${\bf 16}$'s have popped up in other sectors. In this
example, with the minimal $k$-matrix ($k_{21}=k_{31}=k_{41}=1/2$),
sectors $V_2+ V_5+V_7+V_8$,
$V_3+V_5+V_6+V_8$ and $V_4+V_6+V_7+V_8$ yield
${\bf \overline{16}}$'s, resulting in
$0$ net chiral generations.  Changing the $k$-matrix to
$k_{72}=k_{74}=k_{83}=1/2$ we find ${\bf 16}$'s in $V_2+V_5+V_8$,
$V_3+V_6+V_8$ and a ${\bf \overline{16}}$ in $V_4+V_7+V_8$,
for 4 net generations.
For all allowed choices of $k_{ij}$ the number of generations is even.

The appearance of unwanted states when new projections are added is the
complication that makes obtaining three generations so challenging. One
strategy for curing this problem is to include, in the basis
vectors added after the NAHE set,
additional twists for the ``hidden sector'' fermions. Since
Ramond boundary conditions raise the
vacuum energy of a sector (Eq. \ref{lenergy}), this
will have the effect of raising the mass of states from sectors
involving this twist. Such a twist will not, however, be left-right
symmetric, and the constraints of Eqs. \ref{wconstraint} will be
more restrictive. For real fermions with $0,1/2$ boundary conditions,
the number of (real) unpaired $1/2$'s must be equal to 0 mod 8, and the
number of (real) $1/2$'s overlapping between two different vectors must
be 0 mod 4.  A sample solution is
\begin{eqnarray*}
V_7=(0_c\ 000\ 000\ 000\ 011\ 011\ 000&|&
00\ 00\ 00\ 11\ 11\ 00\ 0^{10} 0^3_c 0^4 1^4 1^4 0^4)
\\
V_8= (0_c\ 011\ 000\ 011\ 000\ 011\ 000&|&
11\ 00\ 11\ 00\ 11\ 00\
0^{10} 0^3_c 1^4 1^4 0^4 0^4 ).
\end{eqnarray*}
These vectors remove the unwanted massless states and leave
precisely three ${\bf 16}$'s.
Note the price we had to pay was breaking apart our
``hidden sector'', which was broken down to $SO(4)^4$
(although it can be enhanced by twisted sector gauge bosons).

An alternate
mechanism for breaking the degeneracy is, instead of
breaking the horizontal $U(1)$ currents,  to break
their pairing with right-moving counterparts.  We may add a twist
which is embedded in the internal $U(1)_i$'s, which label the
two ${\bf 16}$'s in a given sector,
but not their
right-moving counterparts.  As this is left-right asymmetric,
the constraints in Eqs. \ref{wconstraint} will be more difficult to
satisfy; in fact, we cannot satisfy them with only $0,1/2$
boundary conditions.
To find solutions, we must include $1/4$ twists, which loosen
the constraints somewhat ($4V_i\cdot V_j=0\ \mbox{mod\ 1}$ instead of
$2V_i\cdot V_j=0\ \mbox{mod\ 1}$).
However, the $V\cdot V$ constraint Eq. \ref{wconstraint},
will restrict the number of $1/4$ boundary conditions
 which can appear.  We must have
\begin{equation}
4\ V\cdot V=  N_{{1\over 2}} +
 {1\over 4}\cdot N_{1\over 4}=0 \quad ({\rm mod}\ 2),
\end{equation}
where $N_{1\over 2}$, $N_{1\over 4}$ are the number of
complex fermions with  $1/2$ and $1/4$ boundary conditions respectively.
If we wish the vector to include 3 asymmetric $1/2$'s, then
the number of $1/4$'s must be an odd multiple of 4.
Thus, letting ``$+$'' represent $V=1/4$,  we  add the vector
\begin{displaymath}
V_7=(0\ (000)^6 | 10\ 01\ 01\ 10\ 10\ 01\ +^5_c
+^3_c +^4_c 0^4_c)
\end{displaymath}
to $V_0-V_6$.
The $1/4$'s  in this vector break the $SO(10)$ gauge group to
$SU(5)$.  In the process, the ${\bf 16}$'s are broken into
their components ${\bf 1}+{\bf 10}+{\bf \bar 5}$ under
$SU(5)$.  The equation determining which states survive
projection is:
\begin{equation}
{1\over 2}N_J+{1\over 4}N_5=
\sum_i k_{7i}\alpha_i \quad (\mbox{mod\ 1})\label{qpro}
\end{equation}
where $N_J$ is the excitation number for the internal $U(1)$
labeling the degenerate ${\bf 16}$'s,
and $N_5$ is the excitation number in the $SU(5)$ sector, equal
to $0$ for the ${\bf 1}$, $2$ for the ${\bf 10}$ and $4$ for
the ${\bf \bar 5}$.  Thus, assuming $\sum_i k_{7i}=0$,
from  ${\bf 16}_\pm$,
a ${\bf 1}+{\bf \bar 5}$ survives from $16_+$ and a
${\bf 10}$ from $16_-$, leaving one complete $SU(5)$ multiplet
from each of three sectors.  Note that to preserve modular invariance,
we were forced to break the hidden sector gauge group also
to $U(4)\times SO(8)$.  This is the mechanism
used to build three generation ``flipped'' $SU(5)$ models
\cite{flipped}.
Note that, in both this approach and the previous one, the
``hidden'' sector fermions played an important role in
allowing us to make the final reduction to three generations.

\section{$SO(10)\times SO(10)$ Models}

To attempt a generalization of the above approaches to a
unified model with
adjoint Higgs representations,
we must do two things: enlarge the gauge group to
$SO(10)\times SO(10)$, and insure that mixed Higgs
representations such as $(\mbox{${\bf 10}$},\mbox{${\bf 10}$})$ actually exist.
The first is easy to
accomplish; we just take a second $SO(10)$ from the $SO(16)$ hidden
sector of the above models.
However, the second feature requires a substantial
modification. This Higgs representation transforms as a vector under two
gauge groups simultaneously; it must be created by exciting two
fermions, one carrying the vector index under each group,  entailing a
large excitation energy. If we limit ourselves to $0,1/2$ boundary
conditions, both fermions must have NS boundary conditions, and the
energy associated with the two fermion excitations will be $+1$.
For the state to be massless, it must therefore
arise from a sector with ground
state energy $E_L=-1$.  The natural candidate for this sector is the
Neveu-Schwarz sector, $V=0$,
with $(E_R,E_L)=(-1/2,-1)$. We will attempt to
arrange our basis vector such that there is a surviving state of the form
\begin{equation}
\bar\chi^I_{-1/2}|0\rangle_R\Psi^A_{-1/2}\Psi^B_{-1/2}|0\rangle_L
\end{equation}
which transforms
as a $(\mbox{${\bf 10}$},\mbox{${\bf 10}$})$ under $SO(10)_A\times SO(10)_B$.

We immediately run into a problem, though, with the set of basis
vectors used in the previous section, the NAHE set.
This set of vectors automatically projects out any states in
the NS sector which would transform under both the original
$SO(10)$ and a group coming from the hidden sector.
Explicitly, the projections must include the sum
\begin{displaymath}
V_0+V_2+V_3+V_4=(0_c 0^{18} | 0^{12} 0^{10} 0^3_c 1^{16}).
\end{displaymath}
This enforces a projection that requires an even number of
fermionic excitations in the $SO(16)$ sector, thus eliminating the
mixed states.  Clearly we must modify this set of vectors.

The fact that mixed representations are projected out of the untwisted
sector is a manifestation of the fact that models based on
the NAHE set are compactifications of the $E_8\times E_8$
string.  Progress can be made by finding a set of vectors
which correspond instead to a compactification of the
$SO(32)$ string\footnote{The advantage of using the $SO(32)$ string
has also been observed in \cite{newfont}.}.
This is accomplished with the set:
\begin{eqnarray}
V_2=(1_c (100)^2 (010)^2 (010)^2&|&
(00)^2(10)^2 (10)^2 1^{10} 0^{10} 0^{10} 1_c)\\
V_3=(1_c (010)^2 (100)^2 (001)^2&|&(10)^2(00)^2(01)^2
0^{10} 1^{10} 0^{10} 1_c)\\
V_P=(0_c (000)^2(000)^2(000)^2&|&(00)^2(00)^2(00)^2 1^{10}
1^{10} 1^{10} 1_c).\label{notnaha}
\end{eqnarray}
$V_0,V_1$ and $V_P$ produce a toroidal compactification of the
$SO(32)$ string (with background fields).
The vectors $V_2,V_3$ provide the $Z_2\times Z_2$
orbifold twists that reduce the $N=4$ supersymmetry to $N=1$,
as was the case for the NAHE set.  However, they differ now
in that each twist is embedded in a different $SO(10)$ factor,
in the process breaking $SO(32)\rightarrow SO(10)^3\times U(1)$.
The vector $V_P$ is just the standard GSO projection of the
ten dimensional $SO(32)$ string.

An examination of the projections in the NS sector now
reveal that the states
\begin{equation}\bar\chi^{1,2}_{-1/2}|0\rangle_R\Psi^{10A}_{-1/2}
\Psi^{10B}_{-1/2}|0\rangle_L,\end{equation}
\begin{equation}\bar\chi^{3,4}_{-1/2}|0\rangle_R\Psi^{10A}_{-1/2}
\Psi^{10C}_{-1/2}|0\rangle_L,\end{equation}
and
\begin{equation}\bar\chi^{5,6}_{-1/2}|0\rangle_R
\Psi^{10B}_{-1/2}\Psi^{10C}_{-1/2}|0\rangle_L\end{equation}
survive, providing
the full set $(\mbox{${\bf 10}$},
\mbox{${\bf 10}$},1)+(\mbox{${\bf 10}$},1,
\mbox{${\bf 10}$})+(1,\mbox{${\bf 10}$},\mbox{${\bf 10}$})$ necessary for
breaking $SO(10)^3\rightarrow SO(10)_D$.

The sectors $V_2$, $V_3$, and $V_0+V_2+V_3$ provide
$({\bf 16},1,1), (1,{\bf 16},1), (1,1,{\bf 16})$ representations
respectively, becoming chiral generations under $SO(10)_D$.
Each is replicated 8 times per sector.
As in the previous section, we may reduce this degeneracy
by adding left-right paired vectors which break the horizontal
symmetries.  Adding
\begin{eqnarray*}
V_4=(0_c\ 011\ 000\ 000\ 000\ 000\ 011 &|&
11\ 00\ 00\ 00\ 00\ 11\ 0^{10} 0^3_c 0^{16}) \\
V_5= (0_c\ 000\ 011\ 011\ 000\ 000\ 000 &|&
00\ 11\ 11\ 00\ 00\ 00\
0^{10} 0^3_c 0^{16} )
\end{eqnarray*}
breaks down the horizontal symmetry to $U(1)^3$, and
leaves a 2-fold degeneracy in each sector associated with the
generational $U(1)$ charge $iy_4y_5$ for $V_2$,
$iy_1\omega_6$ and $i\omega_2\omega_3$ for $V_3$ and
$V_2+V_3+V_0$ respectively.  This again leaves a total of six
chiral generations.
Adding further projections, to reduce
this number, again produces unwanted states in new sectors.
In the previous section, we saw that this could be circumvented
by making proper use of the hidden sector.
However, in these models, with the extended gauge sector,
there is essentially no hidden sector left, and
we cannot apply the tricks used in the previous section.
All examples we have constructed
have an even number of generations.

All our attempts to use the standard $Z_2\times Z_2$
mechanism for generating three
generation models fail, and we must look for an alternate method. We
will take the approach of only using two of the three $Z_2\times
Z_2$ twisted sectors for chiral generations, and not attempting to
treat the different sectors symmetrically ({\it i.e.}, have two
generations associated with one twist, one with the other). We
start with the basic set defined above, producing $SO(10)^3$, with $8$
${\bf 16}$'s per sector. We will add additional vectors which preserve the
first two $SO(10)$'s, and preserve the
$(\mbox{${\bf 10}$},\mbox{${\bf 10}$})$ representations in
the untwisted sector, but we will allow ourselves
the freedom to break apart the
third $SO(10)$.

The number of massless ${\bf 16}$'s coming from a given sector is
$2^{7-n_i}$ where $n_i$ is the number of independent projections
imposed on the zero modes by the generalized GSO
projections (assuming $0,1/2$ boundary conditions).
Since this is a power of two,
if we are going to have an odd number of ${\bf 16}$'s (three), there
must be at least one sector for which $n_i=7$.  This requires
that at least three additional vectors  be added to the above set.

We will attempt to add additional vectors such that sector
$V_3$ has $n=7$ and $V_2$ has $n=6$.
We add, to $V_2,V_3,V_P$, the vectors
\begin{eqnarray*}
V_4=(0_c 011\ 000\ 000\ 000\ 000\ 011 &|&11\ 00\ 00\ 00\ 00\ 11\
0^{10}\ 0^{10} (0^4 0^2)_c) \\
V_5=(0_c 000\ 011\ 011\ 000\ 000\ 000 &|&00\ 11\ 11\ 00\ 00\ 00\
0^{10}\ 0^{10} (1^4 0^2)_c) \\
V_6=(0_c 011\ 000\ 011\ 000\ 000\ 000 &|&11\ 00\ 11\ 00\ 00\ 00\
0^{10}\ 0^{10} (0^4 0^2)_c).
\end{eqnarray*}
With this set of basis vectors, the gauge group has been broken
to $SO(10)^2\times SO(8)\times U(1)^3$. Unfortunately,
the available  ``hidden sector'' has not provided enough freedom
to inhibit additional chiral sectors from appearing.
We find a total of four sectors
for which $n_i=7$; writing them as $V=\sum_i \alpha_i V_i$ these
sectors are
\begin{eqnarray}
\vec\alpha^1&=&(0,0,0,1,0,0,0,0)\nonumber\\
\vec\alpha^2&=&(0,0,0,1,0,1,1,1)\nonumber\\
\vec\alpha^3&=&(0,0,0,1,1,0,0,0)\nonumber\\
\vec\alpha^4&=&(0,0,0,1,1,1,1,1)\label{alphao}
\end{eqnarray}
where
\begin{equation}
\vec\alpha=(\alpha_0,\alpha_1,\alpha_2,\alpha_3,\alpha_4,\alpha_5,
\alpha_6,\alpha_P).
\end{equation}
In addition, there are four sectors for which $n_i=6$,
\begin{eqnarray}
\vec\alpha^5&=&(0,0,1,0,0,0,0,0)\nonumber\\
\vec\alpha^6&=&(0,0,1,0,1,0,1,0)\nonumber\\
\vec\alpha^7&=&(1,1,1,0,0,1,1,0)\nonumber\\
\vec\alpha^8&=&(1,1,1,0,1,1,0,0).\label{alphat}
\end{eqnarray}
We refer to these as being degenerate sectors, because they
produce multiple ${\bf 16}$'s.
In these sectors,
some of the projections on the massless states in these sectors
are redundant. Then, if the $k$-matrix is chosen such that the redundant
projections are incompatible, all  massless states from the sector can
be eliminated. However, eliminating one of these sectors only changes
the net number of chiral ${\bf 16}$'s,
 $N({\bf 16})-N({\bf \overline{16}})$,  by an even
number (two). For the sectors $\vec\alpha^{1-4}$, all projections are
independent, and it is impossible to eliminate the ${\bf 16}$ coming from
this sector by altering the $k$-matrix. One can change the chirality of
the ${\bf 16}$ from one of these sectors, but again, this changes the net
number of $N({\bf 16})-N({\bf \overline{16}})$ by two.
Clearly, if there is to be an odd
number for $N({\bf 16})-N({\bf \overline{16}})$,
the number of sectors for which $n_i=7$
must be odd.

Unfortunately, it seems to be a generic property
of basis vectors within this framework that massless sectors
with $n_i=7$ always occur in sets of 4. We have not been able to
construct examples for which this is not true. Thus, it does not seem
possible to obtain three generation models based on
$SO(10)\times SO(10)$ within our framework.

\section{$SU(5)\times SU(5)$ Models}

We will attempt to get around this obstruction by
breaking $SO(10)^2\rightarrow SU(5)^2$. We note that
if the vector $V_P$ is
converted to a $1/4$ moded vector, breaking $SO(10)\rightarrow SU(5)$,
the projection acting on a ${\bf 16}$ will not leave a complete multiplet.
If we replace $V_P$ with
\begin{equation}
V_7=(0_c 0^{18}| 0^{12}\ +^{16}_c)\label{break}
\end{equation}
($V_P=2V_7$, so all the above sectors are still here),
the projection will be of the form
\begin{equation}
{1\over 4} N_5 = \sum_i \alpha_i k_{7i}\quad (\mbox{mod\ 1}).
\end{equation}
Depending on whether the right hand side equals $1/2$ or $0$,
either a \mbox{${\bf 10}$}\ or $1+\mbox{${\bf\bar5}$}$ will survive.
Of course, both a \mbox{${\bf 10}$}\ and
a \mbox{${\bf\bar5}$}\ are needed to make
a complete family.
Since string theory does not have nonabelian gauge anomalies, if a
\mbox{${\bf 10}$}\ arises in one sector,
 a \mbox{${\bf\bar5}$}\ (or \mbox{${\bf \overline{10}}$})
 must show up somewhere
else\footnote{It is also possible that $(\mbox{${\bf 5}$},
\mbox{${\bf 5}$})$ representations
would appear as part of the anomaly cancellation. These would transform
as $\mbox{${\bf 10}$}+{\bf 15}$'s under $SU(5)_D$, which is unacceptable
phenomenologically.  However, the appearance of these representations
is easily prevented by an appropriate choice of the $k$-matrix.}.
In practice, it is easier to keep track of
\mbox{${\bf 10}$}'s; because two oscillator excitations
are required to form a \mbox{${\bf 10}$},
these states generally only come
from sectors with Ramond boundary conditions
in the $SU(5)$ sector; {\it i.e.} the ${\bf 16}$ sectors of the
original $SO(10)$ model.  Since  only one excitation is needed for a
\mbox{${\bf\bar5}$}, these can come from a
variety of new sectors which have $1/4$
boundary conditions in the $SU(5)$ sector and have $E_L\le -1/4$.

Our strategy then is to construct an even generational
$SO(10)^2$ model, break  $SO(10)^2\rightarrow SU(5)^2$,
and use the freedom of the $k$-matrix to control the number of
\mbox{${\bf 10}$}'s that appear in the original sectors of the $SO(10)$ model,
allowing the string to fill in the remaining \mbox{${\bf\bar5}$}'s, such that
a net total of three chiral families remain.
Given that there are some 29 independent $k$-matrix elements,
it does not seem unreasonable that one could do this.

Let us write down the $k$-matrix equations which determine
whether a \mbox{${\bf\bar5}$}\ or
\mbox{${\bf 10}$}\ survives in a given sector.
We find:
\begin{eqnarray*}
\vec\alpha^1:\quad 0&=&k_{73}\\
\vec\alpha^2:\quad 0&=&k_{73}+k_{75}+k_{76} \\
\vec\alpha^3:\quad 0&=&k_{73}+k_{74}  \\
\vec\alpha^4:\quad 0&=&k_{73}+k_{74}+k_{75}+k_{76}
\end{eqnarray*}
must be satisfied
for a \mbox{${\bf 10}$}\ to survive.
The above equations are valid modulo 1.
Recall that the above $k$-matrix
elements must be $0$ or $1/2$ (see Eq. \ref{kconstraint}).
Unfortunately, we cannot
find any set of $k_{ij}$'s such that an odd number of the
above equations are violated.  This is because the above
equations are not independent; adding the four of them gives
zero \mbox{mod\ 1}.  This is a direct consequence of the fact
that $\sum_1^4 \vec\alpha^i=0\ ({\rm mod}\ 2)$, as can readily
be seen from Eq. \ref{alphao}.  The set of massless sectors with
$n_i=7$ closes under addition.  This fact prevents us from
finding a projection in which an odd number of \mbox{${\bf 10}$}'s survive.
This property holds even if we modify $V_7$ to include additional
$1/2$ boundary conditions for internal fermions.  In all cases,
we are stuck with an even number of generations.

To summarize, we have found that we are unable to break out of even
numbers for $N(\mbox{${\bf 10}$})-N(\mbox{${\bf \overline{10}}$})$,
because various $k$-matrix equations
degenerate, and this in turn follows from the fact that the set of
sectors which produce massless ${\bf 16}$'s, both the nondegenerate and
the degenerate ones, are separately closed under addition.
This property seems to be common to
all sets of basis vectors constructed within the framework
of $SO(10)^2$ models with left-right pairing. We have
not encountered a set of basis vectors based on Eq. \ref{notnaha}\ and
preserving the $(\mbox{${\bf 10}$},\mbox{${\bf 10}$})$
and the left-right pairing which
does not have this property.

However, we have made some progress, in that now we know
exactly what feature to look for in a set of basis vectors
as we enlarge the scope of our search:
we must look for an $SO(10)^2$ model  for which the set of non-degenerate
${\bf 16}$ generating sectors does not close on itself.
We have chosen vectors which maintain a left-right pairing until now,
because it provides a convenient way of finding solutions to Eqs.
\ref{wconstraint}. However, this may be  too restrictive.  Notice that
the triplet constraint, Eq. \ref{triplet},
 imposes severe restrictions on right-hand side boundary conditions,
and these restrictions are carried over to the left-hand side by the
left-right pairing. We will now consider models which are made purely
of complex fermions, with no left-right pairing imposed. We start with
the same set as before, now written in purely complex notation:
\begin{eqnarray*}
V_2=(1\ 100\ 010\ 010&|& 1^5\ 0^5\ 000001000011 ) \\
V_3=(1\ 010\ 100\ 001&|& 0^5\ 1^5\ 000001001100 ) \\
V_P=(0\ 000\ 000\ 000&|& 1^5\ 1^5\ 111111000000 ).
\end{eqnarray*}

We must find three additional vectors which satisfy
Eq. \ref{wconstraint}.  Finding additional independent
solutions of Eq. \ref{wconstraint}
becomes rather tedious as the number of basis vectors
is increased, but can be done by computer.
Here is a sample set of basis vectors which satisfies
Eq. \ref{wconstraint}\
and preserves the $(\mbox{${\bf 10}$},\mbox{${\bf 10}$})$ Higgs:
\begin{eqnarray}
V_4=(0\ 000\ 000\ 000&|& 0^5\ 0^5\ 100111110101 ) \nonumber\\
V_5=(0\ 000\ 011\ 000&|& 0^5\ 0^5\ 000110101110 ) \nonumber\\
V_6=(0\ 000\ 000\ 000&|& 0^5\ 0^5\ 111100001111 ).\nonumber\\
\label{works}
\end{eqnarray}
This set gives the gauge group
$SO(10)^2\times SU(2)^2\times U(1)^{10}$.
The following sectors give massless, non-degenerate ${\bf 16}$'s:
\begin{eqnarray*}
\vec\alpha^1&=&(0,0,1,0,0,0,0,0)\\
\vec\alpha^2&=&(0,0,1,0,0,0,1,1)\\
\vec\alpha^3&=&(0,0,1,0,0,1,1,1)\\
\vec\alpha^4&=&(0,0,1,0,1,1,0,0).
\end{eqnarray*}
The following sectors provide degenerate ${\bf 16}$'s:
\begin{eqnarray*}
\vec\alpha^5&=&(0,0,0,1,0,0,0,0)\\
\vec\alpha^6&=&(0,0,0,1,0,0,1,1),
\end{eqnarray*}
and ${\bf 16}+{\bf \overline{16}}$'s come from:
\begin{eqnarray*}
\vec\alpha^7&=&(1,1,0,1,0,1,1,0)\\
\vec\alpha^8&=&(1,1,0,1,1,1,0,1).
\end{eqnarray*}

Once again, the number of non-degenerate ${\bf 16}$ producing
sectors is even,
and it is impossible  to obtain an odd number of
${\bf 16}$'s in an $SO(10)^2$ model.  This is again seems to be true for
all sets of basis vectors satisfying
our constraints.  However, there is an important difference between
the above set of vectors and those of the previous example.
One readily sees that
\begin{equation}
\sum_1^4 \vec\alpha^i=(0,0,0,0,1,0,0,0)\quad {\rm mod}\ 2. \end{equation}
The massless, nondegenerate sectors do not form a closed set by
themselves, even though it is true that the larger set of massless
sectors is closed;
$\sum_1^8 \vec\alpha^i=0\ {\rm mod}\ 2$.
This removes the obstacle we had
encountered before in finding projections which
left an odd number of \mbox{${\bf 10}$}'s, after breaking $SO(10)$
down to $SU(5)$.

 \begin{table}{{\bf Table 1.}  \mbox{$SU(5)\times SU(5)$}
 Spectrum for Model 1.}\\[1ex]
\begin{tabular}{|l|l||l|l|} \hline
Sector & States & Sector & States\\ \hline\hline
00010000 & \mbox{$2\times$}(\mbox{${\bf 10}$},1)& 00100000 &
(1,\mbox{${\bf 10}$}) \\ \hline
00100003 & (1,\mbox{${\bf\bar5}$}) & 00100012 &
(\mbox{${\bf \overline{10}}$},1) \\ \hline
00100013 & (\mbox{${\bf 5}$},1) & 00100112 & (\mbox{${\bf\bar5}$},1) \\ \hline
00100113 & (1,\mbox{${\bf\bar5}$}) & 00101100 & (1,\mbox{${\bf 10}$}) \\ \hline
00101103 & (1,\mbox{${\bf\bar5}$}) & 01000001 &
(1,\mbox{${\bf\bar5}$})+(\mbox{${\bf\bar5}$},1) \\ \hline
01000003 & (1,\mbox{${\bf 5}$})+(\mbox{${\bf 5}$},1) &
10110110 & (\mbox{${\bf\bar5}$},1)+(1,\mbox{${\bf 5}$})
\\ \hline
10111000 & (\mbox{${\bf\bar5}$},1)+\mbox{$2\times$}
(1,\mbox{${\bf\bar5}$})+(1,\mbox{${\bf 5}$}) &
10111010 &\mbox{$2\times$}(\mbox{${\bf 5}$},1)+
(1,\mbox{${\bf 5}$})+(\mbox{${\bf\bar5}$},1)  \\ \hline
10111110 & (1,\mbox{${\bf 5}$})+(\mbox{${\bf\bar5}$},1) &
10111111 & (1,\mbox{${\bf\bar5}$}) \\ \hline
10111113 & (\mbox{${\bf 5}$},1) &
11010111 & (\mbox{${\bf\bar5}$},1) \\ \hline
11010113 & (\mbox{${\bf 5}$},1) &
11011101 & (\mbox{${\bf\bar5}$},1) \\ \hline
11011102 & (\mbox{${\bf 10}$},1)+(\mbox{${\bf \overline{10}}$},1) &
11011103 & (\mbox{${\bf 5}$},1)  \\ \hline
00000000 & (\mbox{${\bf
5}$},\mbox{${\bf\bar5}$})+(\mbox{${\bf\bar5}$},
\mbox{${\bf 5}$}) & & \\ \hline
\end{tabular}
\end{table}

Changing $V_P$ to $V_7$ as in Eq. (\ref{break}),
\begin{equation}
V_7=(0^{10}|\ +^{10}\ ++++++000000)
\end{equation}
the complete set of basis vectors is now
\begin{eqnarray}
V_0 = (1\ 111\ 111\ 111 &|& 1^5\ 1^5\ 111111111111) \nonumber\\
V_1 = (1\ 100\ 100\ 100 &|& 0^5\ 0^5\ 000000000000) \nonumber\\
V_2 = (1\ 100\ 010\ 010 &|& 1^5\ 0^5\ 000001000011) \nonumber\\
V_3 = (1\ 010\ 100\ 001 &|& 0^5\ 1^5\ 000001001100) \nonumber\\
V_4 = (0\ 000\ 000\ 000 &|& 0^5\ 0^5\ 100111110101) \nonumber\\
V_5 = (0\ 000\ 011\ 000 &|& 0^5\ 0^5\ 000110101110) \nonumber\\
V_6 = (0\ 000\ 000\ 000 &|& 0^5\ 0^5\ 111100001111) \nonumber\\
V_7 = (0\ 000\ 000\ 000 &|& +^5\ +^5\ ++++++000000).\label{cset}
\end{eqnarray}
The $k$-matrix equations determining whether
 \mbox{${\bf 10}$}'s or \mbox{${\bf\bar5}$}'s survive
in $\vec\alpha^1-\vec\alpha^4$ are:
\begin{eqnarray}
\vec\alpha^1:\quad 0&=&k_{72}\nonumber\\
\vec\alpha^2:\quad 0&=&k_{72}+k_{76}\nonumber\\
\vec\alpha^3:\quad 0&=&k_{72}+k_{75}+k_{76}\nonumber\\
\vec\alpha^4:\quad 0&=&k_{72}+k_{74}+k_{75}.\label{cansolve}
\end{eqnarray}
It is now straightforward to find solutions which
eliminate \mbox{${\bf 10}$}'s from
any sector we choose.  For example, the choice $k_{74}=k_{75}=1/2$
(along with $k_{21}=k_{31}=1/2$ to preserve $N=1$ SUSY)
eliminates \mbox{${\bf 10}$}'s from $\vec\alpha^3$ while
keeping them in the other
three sectors.  Actually, for this particular
$k$-matrix, we find $(1,\mbox{${\bf 10}$})$'s in
$\vec\alpha^1, \vec\alpha^4$ and
a $(\mbox{${\bf \overline{10}}$},1)$ in $\vec\alpha^2$, as
well as $2\times (\mbox{${\bf 10}$},1)$
from $\vec\alpha^5$ and $2\times (1,\mbox{${\bf 10}$})$
from $\vec\alpha^6$, for a total
$N(\mbox{${\bf 10}$})-N(\mbox{${\bf \overline{10}}$})=5$. This is easily
fixed by setting $k_{43}=1/2$, to eliminate $\vec\alpha^6$;
the nonvanishing lower half diagonal $k$-matrix elements for this
model are then
\begin{equation}
k_{21}=k_{31}=k_{43}=k_{74}=k_{75}={1\over 2}.\end{equation}
The net
total for $N(\mbox{${\bf 10}$})-N(\mbox{${\bf \overline{10}}$})$ in this
model is now three ($3\times \mbox{${\bf 10}$} +
\mbox{${\bf 10}$}+\mbox{${\bf \overline{10}}$}$).
Computing the complete massless spectrum for this model,
we indeed find that the correct number of
\mbox{${\bf\bar5}$}'s do appear in other
sectors to cancel both $SU(5)$ anomalies.  The total gauge group with
this particular $k$-matrix is $SU(5)\times SU(5)\times SU(4)\times
SU(2)\times U(1)^{10}$ (the hidden sector gauge group is enhanced by
twisted sector gauge bosons). In addition to the above states, there
are an additional $12\times(\mbox{${\bf 5}$}+\mbox{${\bf\bar5}$})$'s,
$47$ singlets
charged under the $U(1)$'s, and
a variety of representations transforming under the hidden sector gauge
group, from various sectors.  The complete
spectrum of $SU(5)\times SU(5)$
nonsinglet states
is shown in Table 1.
 \begin{table}{{\bf Table 2.}  \mbox{$SU(5)\times SU(5)$}
 Spectrum for Model 2.}\\[1ex]
\begin{tabular}{|l|l||l|l|} \hline
Sector & States & Sector & States\\ \hline\hline
00010003 & \mbox{$2\times$}(\mbox{${\bf\bar5}$},1)& 00010011 &
\mbox{$2\times$}(1,\mbox{${\bf\bar5}$}) \\ \hline
00010012 & \mbox{$2\times$}(1,\mbox{${\bf\bar5}$}) & 00100000 &
(1,\mbox{${\bf 10}$}) \\ \hline
00100003 & (\mbox{${\bf 5}$},1) & 00100011 & (1,\mbox{${\bf 5}$}) \\ \hline
00100012 & (\mbox{${\bf 10}$},1) & 00100111 & (\mbox{${\bf\bar5}$},1)\\ \hline
00101100 & (1,\mbox{${\bf 10}$}) & 00101103 &
\mbox{$2\times$}(\mbox{${\bf 5}$},1) \\ \hline
10110110 & (1,\mbox{${\bf\bar5}$})+(\mbox{${\bf 5}$},1) &
10110111 & (\mbox{${\bf\bar5}$},1)
\\ \hline
10110113 & (1,\mbox{${\bf 5}$}) &
10111000 & (\mbox{${\bf 5}$},1)+(\mbox{${\bf\bar5}$},1)+
(1,\mbox{${\bf 5}$})+(1,\mbox{${\bf\bar5}$}) \\ \hline
10111011 & (1,\mbox{${\bf\bar5}$})+(\mbox{${\bf\bar5}$},1)&
10111013 & (1,\mbox{${\bf 5}$})+(\mbox{${\bf 5}$},1) \\ \hline
10111110 & (\mbox{${\bf\bar5}$},1)+(1,\mbox{${\bf 5}$}) &
11010110 & (1,\mbox{${\bf 5}$})+(1,\mbox{${\bf\bar5}$}) \\ \hline
00000000 & (\mbox{${\bf 5}$},\mbox{${\bf\bar5}$})+
(\mbox{${\bf\bar5}$},\mbox{${\bf 5}$}) & &  \\ \hline
\end{tabular}
\end{table}

With this set of basis vectors (\ref{cset}),
we can now use the $k$-matrix to tailor the
appearance of \mbox{${\bf 10}$}'s more or
less to our desires.  For example, we can eliminate \mbox{${\bf 10}$}'s coming
from the degenerate sectors by choosing the $k$'s to change them to
\mbox{${\bf 5}$}'s, and then alter the choice of $k$'s to switch the
\mbox{${\bf \overline{10}}$}\ in  $\vec\alpha^2$ to a \mbox{${\bf 10}$}.
Choosing
\begin{equation}
k_{21}=k_{31}=k_{63}=k_{65}=k_{70}=k_{73}=k_{74}=k_{75}=1/2,
\end{equation}
we indeed find \mbox{${\bf 10}$}'s
from $\vec\alpha^{1,2,4}$ only, with the rest of
the spectrum very similar to the previous example. The $SU(5)\times
SU(5)$ nonsinglet spectrum for this model is shown in Table 2.
To give a further
example of the kind of freedom we have, suppose we
wanted to have a model in  which some sectors contained complete
$1+\mbox{${\bf\bar5}$}+\mbox{${\bf 10}$}$ multiplets.
This can be accomplished by adding $1/2$
twists in $V_7$ such that complete multiplets survive from
$\vec\alpha^{5,6}$, in the same manner as in Eq. \ref{qpro}.  A careful
examination of these sectors suggest the choice
\begin{equation}
V_7=(0\ 011\ 000\ 000\ | +^{10}\ ++++++000000). \end{equation}
Along with a careful choice of $k$-matrix:
\begin{equation}
k_{21}=k_{31}=k_{72}=k_{73}=k_{74}=k_{76}=1/2 \end{equation}
we find $1+\mbox{${\bf\bar5}$}+\mbox{${\bf 10}$}$'s in sectors $\vec\alpha^5$,
$\vec\alpha^6$, and a \mbox{${\bf 10}$}\ in sector $\vec\alpha^4$, with
\mbox{${\bf\bar5}$}'s and \mbox{${\bf 5}$}'s
arising in various additional sectors,
for a total of exactly three generations plus
14 $\mbox{${\bf 5}$}+\mbox{${\bf\bar5}$}$ pairs.

The set of basis vectors, Eq. \ref{works}, is not a unique
set for which this method works.  A computer search reveals
a large number of solutions with the required property,
that the non-degenerate massless sectors not form a closed
set.  It is not clear how many are genuinely distinct and
how many are just relabellings and permutations of the others,
but at least some are distinct because they contain numbers
of degenerate massless sectors.
For example, making the replacement
\begin{equation}
V_6=(0\ 000\ 000\ 011\ |\ 0^5\ 0^5\ 101101000011) \end{equation}
in Eq. \ref{works}\
leads to a set of basis vectors which has four non-degenerate
${\bf 16}$ sectors and only two sectors with degenerate ${\bf 16}$'s,
while the choice
\begin{equation}
V_6=(0\ 011\ 011\ 000\ |\ 1^5\ 1^5\ 110000110011) \end{equation}
leads to a model with six sectors providing degenerate ${\bf 16}$'s.

Much work needs to be done to see if these models are viable
phenomenologically.
These models all posess a large number of unwanted
particles (14 $\mbox{${\bf 5}$}+\mbox{${\bf\bar5}$}$
pairs and $50$ or so singlets).
Since these are in real $SU(5)\times SU(5)$ representations
there is no obstacle in principle to their getting large masses
after some moduli are given vacuum expectation values, although
to actually see if this happens would require a detailed
analysis of F and D-flat directions in these models.
A preliminary look indicates that it would be very difficult
to obtain a single large Yukawa coupling for the top quark in  these
models, and the question of dimension four operators
baryon number violating operators is not clear.
The most serious problem in supersymmetric
grand unified theories, the doublet-triplet splitting for
the Higgs bosons,
has not been addressed.

\section{Conclusions}

Within the free fermion constructions, we have seen that it
is surprisingly difficult, if not impossible, to construct
three generation models based on the $SO(10)$ group.
This mirrors the results of other groups, pursuing
different approaches.  Whether there is any deep signifigance
to this, or it is simply an artifact of the free fermion
construction, is not clear.

We have, however,
identified a way of constructing three generation models
based on $SU(5)$.  The number three does not arise naturally
in these models, but arises only after a careful choice of
projections.  While the specific models constructed here
do not appear to have any obvious phenomenological virtues, beyond
the number of generations, they do demonstrate that there is
at least one way around a barrier that has faced string
GUT model building.  It will be interesting to see if these
constructions can be turned into realistic grand unified
theories.
\bigskip

\centerline{\large Acknowledgements}

\noindent
It is a pleasure to thank M. Peskin for his encouragement and
suggestions, L. Dixon for reading an early version of this
manuscript, and J. Feng for participation in the early stages
of this work.
This work was supported in part by the Department
of Energy under grant \#DE-AC03-76SF00515.

\bigskip
\noindent
Note Added:  After the completion of this work,
\cite{afiu} appeared, which addresses some related issues from the
orbifold approach.

\end{document}